\documentclass[preprint,aps]{revtex4}
\usepackage{amsmath}

\begin{document}

\title{Zitterbewegung is not an observable}

\author{R. F. O'Connell}
\affiliation{Department of Physics and Astronomy, Louisiana State
University, Baton Rouge, LA 70803-4001}
\date{\today}

\begin{abstract}
It has recently been claimed that Zitterbewegung has been observed.  However, we argue that it is not an observable and that the authors' observations must be reinterpreted.
\end{abstract}

\maketitle

The Dirac equation for a free particle \cite{sakurai67} leads to the result that, for an eigenstate of momentum and energy, the velocity is not simply proportional to the momentum but also contains a second term which fluctuates rapidly.  As a result, the coordinate as a function of time exhibits not only the usual uniform rectilinear motion but also contains an additional term, corresponding to very rapid oscillations (dubbed Zitterbewegung by Schr\"{o}dinger in 1930).  It was recently claimed \cite{gerritsma10} that this motion has now been measured by performing a quantum simulation of the one-dimensional Dirac equation using a single trapped ion to behave as a free relativistic quantum particle.  Here, we argue against this claim by pointing out that Zitterbewegung is not an observable.  The key reason for the appearance of this term is the fact that, whereas momentum $\vec{p}$ is a constant of the motion in the Dirac equation, velocity $\vec{v}$ is not.  However, this is simply a consequence of the representation of the Dirac matrices and one can choose other representations (where $\vec{p}\sim\vec{v}$) which are equally valid but do not exhibit this strange motion \cite{oconnell10}.

The resolution of this apparent problem goes back to the work of Moller \cite{moller72}, who pointed out that a body with spin $\vec{S}$ has a \underline{minimum radius} equal to $\vec{S}/mc$.  In addition, in special relativity, one has to generalize $\vec{S}$ to an anti-symmetric tensor $S_{\alpha\beta}=-S_{\beta\alpha}$, which reduces to the 3-vector $\vec{S}$ in the rest-frame of the particle.  In order to accomplish this it is necessary to define a spin supplementary condition  $v^{\alpha}S_{\alpha}=0$ (corresponding to $\vec{v}=0$ in the rest frame) or  $p^{\alpha}S_{\alpha}=0$ (corresponding to $\vec{p}=0$ in the rest frame) or combinations thereof.  In essence, the choice of a spin supplementary condition is equivalent to choosing a new center-of-mass for the spinning particle so that the initial coordinate is charged by a small quantity $\leq\vec{S}/mc$ (which is $\leq 10^{-11}cm$ for elementary particles), which automatically leads to a new momentum-velocity relationship.  Arranging for $\vec{p}\sim\vec{v}$ is the simplest choice of representation.  This concept was discussed in detail by Barker and the present author \cite{barker74} in the context of general relativity, on work relating to spin and orbital precession of a gyroscope, where an apparent discrepancy with the work of Schiff was resolved and where it was shown that all observable results are independent of the choice of coordinates or, concomitantly, the choice of spin supplementary conditions.

Turning now to relativistic quantum mechanics, the simplest dynamics arises from arranging for $\vec{p}$ and $m\vec{v}$ to be proportional.  In fact, it was pointed out by the present author and Wigner \cite{oconnell77} that this is possible for all localized states and arbitrary spin.  However, in the case of the Dirac particle, it was demonstrated in detail by Foldy and Wouthuysen \cite{foldy50}.  In particular, the latter authors changed the position operator by adding terms involving the spin matrices and they also made a corresponding change in the spin operators, (in order to ensure that the correct commutation relations still held) the net result for a free particle being a representation in which the momentum and velocity are proportional, with no Zitterbewegung appearing.  We should emphasize that the Foldy-Wouthuysen transform is now regarded as an integral part of the discussion of the relativistic theory of the electron.  In fact, the well-known book by Bjorken and Drell \cite{bjorken} devotes a whole chapter to it.  In particular, these authors state that "--- it is instructive to cast the Dirac theory in a form which displays the different interaction terms --- in a --- easily interpretable form", by constructing a unitary transformation (the Foldy-Wouthuysen) which, of course, does not change the physics.  However, it should be noted that energy level terms are the only observables, not expectation values of position operators.

The conclusion is that Zitterbewegung is not an observable.  The fact that the authors of \cite{gerritsma10} choose to analyze the motion of trapped ions in terms of a one dimensional Dirac equation does not alter the fact that the ions still possess spin and it does not mitigate our general argument, the essence of which is the use of a coordinate transformation.  Thus, we feel that the results of ref. 2 must be reinterpreted.  It might be thought that experimentalists can select a particular representation but experiments do not distinguish between equally valid but different representations used by theoreticians and which all lead to the same observables.

Turning to the interpretation of the experimental results, the authors of \cite{gerritsma10} state that, in order to study Zitterbewegung, "- - it is necessary to measure $\langle\hat{x}(t)\rangle$, the expectation value of the position operator" and they proceed by stating that "- - the data were fitted with a heuristic model function - -", which they then use to extract information which they claim is Zitterbewegung.  However, the fact is that, even for a non-relativistic non-spinning quantum particle, $\langle\hat{x}(t)\rangle$ is not an observable since, because of the coordinate momentum uncertainty principle, it has uncertainties associated with it.  Thus, in our view, the authors are simply measuring the time-dependent fluctuations in $\langle\hat{x}(t)\rangle$.

\section*{Acknowledgements} This work was partially supported by the National Science Foundation under Grant No. ECCS-0757204.

\end{document}